\documentclass[11pt,a4paper]{article}

\pdfoutput=1

\newcommand{\beq}{\begin{equation}}

\newcommand{\eeq}{\end{equation}}

\def\be{\begin{equation}}
\def\ee{\end{equation}}
\def\beqa{\begin{eqnarray}}
\def\eeqa{\end{eqnarray}}

\usepackage[pdftex]{graphicx}
\usepackage{jcappub}

\title{Watchers of the multiverse}

\author[a,b]{Jaume Garriga} 
\author[c]{Alexander Vilenkin}

\affiliation[a]{Departament de Fisica Fonamental i Institut de Ciencies del Cosmos,
\\ Universitat de Barcelona, Marti i Franques 1, 08028 Barcelona, Spain}

\affiliation[b]{ Yukawa Institute for Theoretical Physics, Kyoto University,
Kyoto 606-8502, Japan}

\affiliation[c]{Institute of Cosmology, Department of Physics and Astronomy,\\ 
Tufts University, Medford, MA 02155, USA}

\abstract{
An unresolved question in inflationary cosmology is the assignment of 
probabilities to different types of events that can occur in the eternally inflating multiverse.
We explore the possibility that the resolution of this ``measure problem" may rely on non-standard dynamics 
in regions of high curvature. In particular, ``big crunch" singularities at the future boundary of  bubbles with 
negative vacuum energy density may lead to bounces, where contraction is replaced by inflationary expansion driven by different vacua in the landscape.  
Similarly, singularities inside of black holes might be gateways to other inflating vacua.
This would drastically affect the global structure of the inflating multiverse. We consider a measure based on a probe geodesic which undergoes an infinite number
of passages through crunches. This can be thought of as the world-line of an eternal ``watcher",  collecting data in an orderly fashion. We compare this to previous approaches to the measure problem. 
The watcher's measure is independent of initial conditions and does not suffer from ambiguities associated with the choice of a cut-off surface. 
Another potential benefit from passing through crunches is that the
observations collected by the watcher may easily depart from
ergodicity, in very generic landscapes. This may significantly
alleviate the problem of Boltzmann Brain dominance. }

\begin{document}

\maketitle

\section{Introduction}

A serious challenge to inflationary cosmology is the problem of
assigning probabilities to different observations,
known as the measure problem. Inflation is generically eternal to the future, so any observation having a nonzero probability occurs an
infinite number of times.  The
relative probability of outcomes $A$ and $B$ resulting from some
measurement can be defined as
\beq
\frac{p_A}{p_B}=\frac{N_A}{N_B}, 
\label{defp}
\eeq
where $N_A$ and $N_B$ are the corresponding numbers of instances.
In the multiverse context, $A$ and $B$ can refer to different values
of some low-energy constants, measured by observers living in
different vacua of the particle physics landscape.  

Both $N_A$ and $N_B$ are infinite in an eternally inflating universe, 
so Eq.~(\ref{defp}) requires a cutoff.  
Most of the measure prescriptions discussed so far involve geometric
cutoffs: the ratio $N_A/N_B$ is evaluated in a finite region of spacetime,
and then the limit is taken when the size of the region goes to
infinity.  The problem is that the result is sensitively dependent on
the limiting procedure. (For an up to date review of the measure
problem, see, e.g., \cite{Freivogel}.)

The simplest measure prescriptions are the
global time cutoffs, where one counts only observations that occurred
prior to some time, $t<t_c$, and then takes the limit $t_c\to\infty$.
An attractive property of these measures is that the resulting
probability distributions do not depend on the choice of the comoving
region that is being sampled, reflecting the attractor behavior of
eternal inflation.
One finds, however, that they do depend on one's
choice of the global time variable $t$ \cite{lime1,lime2,GarciaBellido}.  A variety of choices have been
considered, e.g., proper time
\cite{lime1,lime2,GarciaBellido,Vilenkin:1994ua}, scale factor
\cite{lime1,lime2,GarciaBellido,DGSV,Bousso:2008hz,DGLNSV}, 
comoving horizon (or `lightcone time') \cite{GSVW,RBLC}, and comoving apparent 
horizon (CAH) \cite{AVCAH,SVCAH}, along with more complicated
prescriptions \cite{GSVW,stationary}. 
Apart from this lack of uniqueness, there are also problems of a more  
technical character. Geodesic congruences that are usually used to
define global time tend to develop caustics; then the time variable
$t$ becomes multi-valued.  Moreover, some of the global time variables
are not generally monotonic, and one needs to introduce additional rules 
to handle these cases.

Another class of measures includes the so-called local measures, which
sample a spacetime region  
in the vicinity of a given timelike geodesic.  Here again, there are a
number of possible choices for the sampling region.  It could be the
past light cone of the geodesic (the causal patch measure \cite{causalpatch}),
the region bounded by the apparent horizon \cite{AHbousso}, or the region
within a fixed physical distance of the geodesic (the `fat geodesic'
measure \cite{Bousso:2008hz}).  

A related proposal, closer in spirit to the one we shall explore here, is that instead of counting
observations made by all observers within a spacetime region 
defined by a geometric cutoff, we include only observations made by a
single `observer' specified by a timelike geodesic.  A simple version of such 
a measure was introduced in \cite{GV98} and later
discussed in \cite{VV,Vanchurin}.  Most recently, the single observer picture was
discussed   
by Nomura \cite{Nomura}, who motivated it from quantum mechanical considerations. 

The basic problem of all local measures, including the single observer measure, 
was pointed out already in \cite{GV98}.
A typical geodesic, starting in some inflating de Sitter (dS) vacuum, 
will traverse a number of dS bubbles and
will eventually enter a terminal bubble -- either an anti-de Sitter
(AdS) bubble terminating at a big crunch, or a bubble of
supersymmetric stable Minkowski vacuum.  All geodesics, except for a set of measure zero,
will visit a finite number  of
bubbles, so the resulting probability distribution will depend on what
geodesic we choose.  Hence, one needs to consider
an ensemble of geodesics with different initial conditions.  Without
specifying such an ensemble, these measures remain essentially
undefined. 

Much of the recent work on the measure problem has been aimed at 
exploring phenomenological aspects of different measure proposals, making sure they are not riddled
with internal inconsistencies or obvious conflict with the data.  Although some
of the measure candidates have already been ruled out in this way,
it seems unlikely that this kind of phenomenological
analysis will yield a unique prescription for the measure.

A more satisfactory approach would be to motivate the choice of
measure from some fundamental theory.  In this spirit, it was proposed
in \cite{GVhm,GVhmci} that the dynamics of the inflationary multiverse has a
dual description in the form of a lower-dimensional Euclidean theory
defined on the future boundary of spacetime.  The measure of the
multiverse can then be related to the short-distance cutoff in that
theory.  This idea has been further explored in
\cite{Bousso:2010id,AVCAH,RBLC}, and the
relation of the resulting measure to geometric cutoff prescriptions
has been investigated in \cite{SVCAH}.  This approach, however, encounters
a serious difficulty with bubbles of negative-energy (AdS) vacua, which
develop big crunch singularities in their interiors in a finite proper
time.  The proposal of \cite{GVhm,GVhmci} was that such bubbles should be
excised from the future infinity, with their interiors being
represented by $2D$ Euclidean theories living on the boundaries of the
excised regions.  Some support for this conjecture came from the
recent work \cite{JM,HS,JG,KSS}.  However, it seems to follow from this work that
a $2D$ boundary theory can give only an approximate description of the
bulk, and the approximation gets very poor in cases when there is a
significant amount of slow roll inflation inside the bubble.

Thus, despite a considerable effort, the measure problem remains
unresolved.  This suggests that some important element may be missing
in our understanding of the multiverse.  Here, we explore the
possibility that the global structure of the multiverse may
significantly differ from what is usually assumed.  Specifically, we
conjecture that spacetime singularities will eventually be resolved 
in the fundamental
theory of Nature, so that the big crunches that occur in AdS bubbles 
will turn out to be nonsingular.  The
standard description of AdS regions would then be applicable at the
initial stages of the collapse, but when the density and/or curvature
get sufficiently high, the dynamics would change, resulting in a
bounce.  Scenarios of this sort have been discussed in the 1980's in
the context of the so-called maximum curvature hypothesis
\cite{FMM1,FMM2}, and more recently in the context of pre-big-bang scenario \cite{Veneziano},  ekpyrotic and cyclic models \cite{ekpyrotic,cyclic}, loop quantum cosmology \cite{Ashtekar,Bojowald} and holographic ideas \cite{Brustein}.  The high energy densities reached near the bounce would trigger
transitions to other vacua of the landscape.  The only terminal vacua
in this picture are stable Minkowski vacua. A very different
framework where the universe can survive AdS crunches has been recently suggested by
Nomura \cite{Nomura2}.

We shall argue that this global structure allows for an improvement in the
definition of the measure.  To explain the idea, 
let us first assume that the landscape does not include stable
Minkowski vacua.  Then, all future-directed timelike 
geodesics would pass through a succession of dS and AdS vacua,
extending all the way to future infinity.  Given a finite number of
vacua in the landscape, a generic geodesic will pass through each
vacuum an infinite number of times.  We can use such a geodesic to
define a local measure -- e.g., causal patch, apparent horizon, or fat
geodesic measure.  One can expect that the
corresponding probability  distributions will not depend on the choice
of a geodesic, except for a set of geodesics of measure zero.  Hence,
there is no need to introduce an ensemble of geodesics.  
We shall refer to this class of measures as `eternal geodesic measures'.

If stable Minkowski vacua do exist, we can still
define a measure by focussing on eternal geodesics that do not get captured in such
vacua.  This measure would assign zero probability to observations
performed in supersymmetric Minkowski vacua.  Such vacua are predicted
to exist in superstring theory, but it appears that they cannot
support nontrivial chemistry and thus are not likely to host observers.\footnote{We note that our measure proposal is in some sense opposite to the census taker measure \cite{census}, which is focussed on eternal observers in terminal Minkowski vacua.  The phenomenology of the census taker measure has not yet been studied.  Some problems with this measure have been pointed out in Ref.~\cite{Freivogel}.} 

As mentioned above, eternal geodesics eliminate the need for
  specifying an ensemble of geodesics. Although this is an improvement
  over the standard approach (where geodesics in the ensemble  
terminate at singularities or at time-like infinity of Minkowski regions),
we should still specify how these geodesics are to be used in order to
count events. Local measures suffer from 
the ambiguity associated with the choice of the sampling region in the vicinity of the geodesic.
 Moreover, none of the current proposals for defining a local
measure is completely satisfactory. The probability distributions for the
cosmological constant $\Lambda$ derived from causal patch and apparent
horizon  
measures have non-integrable divergences at $\Lambda\to 0$ \cite{Salem,BFLR}.  The
divergence is particularly strong at $\Lambda\to -0$, and thus these
measures make a strong prediction that $\Lambda$ should have a very
small negative value.  This is of course in conflict with observation.
The fat geodesic measure does not have this problem, but its
implementation encounters other difficulties, which will be
discussed in the next Section. 

In the present paper, we shall adopt a different strategy, where
  extended events (or ``stories" \cite{guthvanchurin}) are counted if
  they are pierced through by the eternal geodesic. 
The details of this measure prescription will be discussed in the
following Section, starting with the case of a landscape with no
terminal vacua.  In Section 3 we set up some formalism necessary for
the calculation of probabilities in this measure.  Section 4 deals
with the question of the arrow of time: whether or not it can exist in
the absence of terminal vacua.  Extensions to landscapes with terminal
Minkowski and AdS vacua and some problems associated with black hole
nucleation are discussed in Sections 5 and 6, respectively.  Finally,
our conclusions are summarized in Section 7. 



\section{Defining probabilities}

\subsection{The watcher measure}

\begin{figure}
\begin{center}
\vspace{-2cm}
\includegraphics[width=14cm]{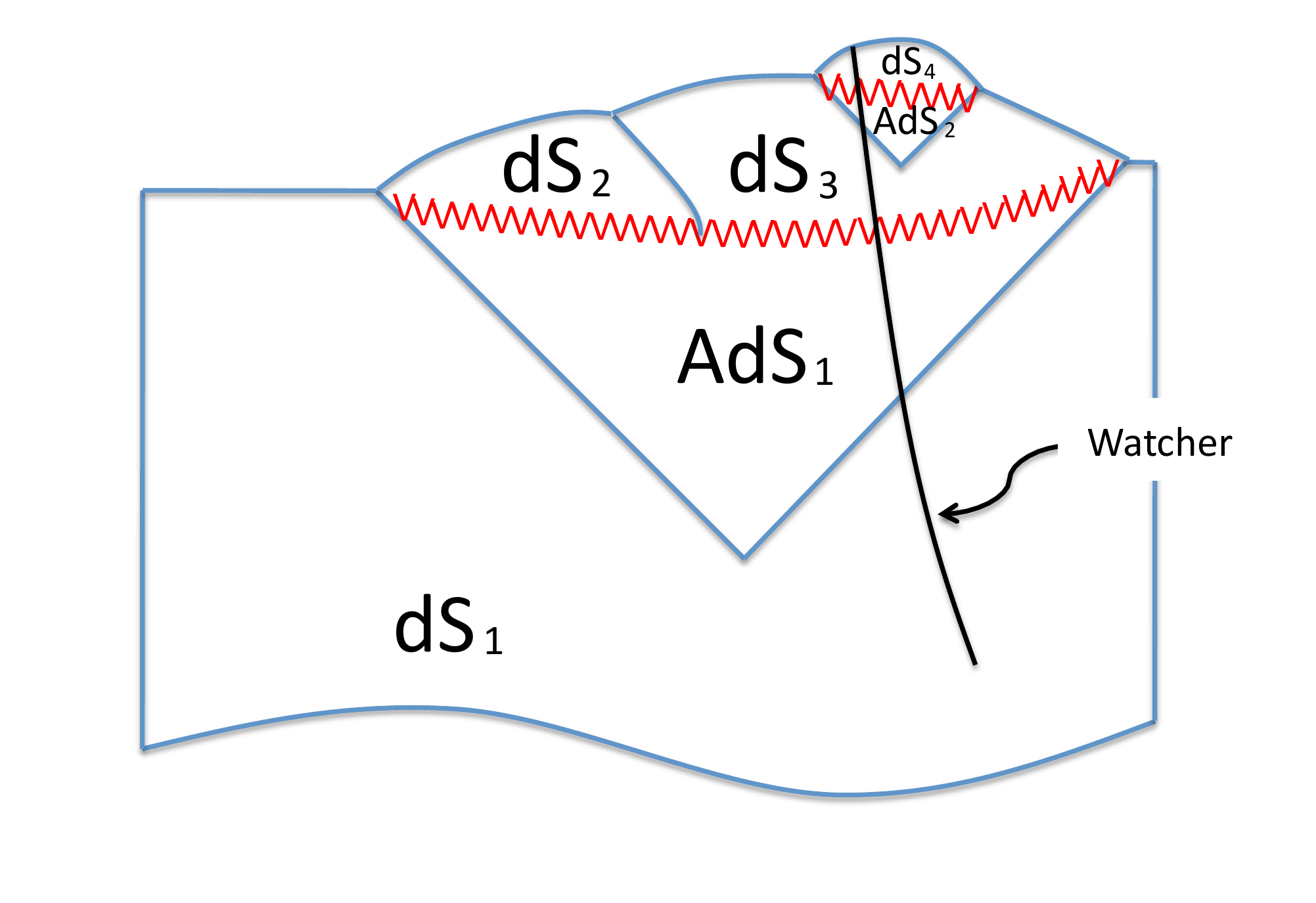}
\vspace{-1.5cm}
\caption{Causal diagram of a multiverse with both positive (dS) and negative (AdS) energy density regions. 
The worldline of the watcher goes through an infinite sequence of AdS crunches.
 \label{watcher}}
\end{center}
\end{figure}

We shall consider an eternally inflating universe populated by regions
of different vacua, with both positive (dS) and negative (AdS) energy
density (see Fig. \ref{watcher}).  To simplify the analysis, we shall first assume that the
landscape does not include 
any Minkowski vacua.  Extension to a more general case will be
discussed later in Section 5.

Our key assumption is that AdS crunches are nonsingular and are
followed by a bounce, so that geodesics can be continued through the
crunch.  
Because of the high energy densities reached near
the bounce, the crunch regions are likely to be excited above the
energy barriers between different vacua, so transitions to other vacua
are likely to occur.  We shall make no assumptions about the dynamics
of the bounce and simply characterize AdS vacua by the
transition probabilities to new (dS or AdS) vacua after the crunch.
Different parts of the same crunch region can, of course, transit to
different vacua.  

We shall assume that the vacuum landscape is irreducible, i.e. any
vacuum can be reached by a sequence of transitions from any other vacuum. 
Given a finite number of vacua in the landscape, a generic timelike
geodesic will then pass through each vacuum an infinite number of times.   
The number of distinguishable events that can be detected by any eternal observer is finite, and thus any event that has a nonzero probability will be detected an infinite number of times \cite{manyworlds}.  
The relative probability of two events, $A$ and $B$, can then be identified with the relative frequency at which the events are encountered along the geodesic.  To avoid confusion 
between the eternal observer and physical observers in the multiverse, from now on we shall refer to the eternal observer as "the watcher".

We shall now spell out what exactly we mean by events "encountered" by a geodesic.  "Events" in General Relativity are often represented by points in spacetime.  However, macroscopic 
events that are of interest to us are extended in both space and time.  Hence, we shall assume that each type of event $A$ is characterized by a finite spacetime domain $D_A$, so it presents 
a certain cross-section $\sigma_A$ for the watcher's geodesic.  
The picture of events as extended entities in spacetime is essentially the same as the notion of "stories" introduced by Guth and Vanchurin in Ref.~\cite{guthvanchurin}.  They define a "story" as "a description of a finite-sized region of spacetime that is specified with well defined tolerances, so that if anybody looked at what was happening in a region of spacetime, she could decide without ambiguity whether or not this story occurs in the region."  We can define the domain $D_A$ as the minimal spacetime region that is necessary to specify the event (story) $A$.
We could then count only events whose domain is traversed by the geodesic.

As it stands, this prescription is not quite satisfactory, since it gives preference to events with a large cross-section.  For example, a measurement that uses bulky equipment or takes a large amount of time will be assigned a higher probability.  In order to correct for this effect, we shall introduce the corrected number of encounters $N_A$,
\beq
N_A = \frac{\sigma_0}{\sigma_A} \nu_A, \label{corrected}
\eeq
where $\nu_A$ is the number of passages through domains of type $A$ and $\sigma_0$ is an arbitrary constant.\footnote{If transdimensional transitions are allowed in the landscape
(see e.g. \cite{transtufts,transcal} and references therein), then the cross sections are defined to be the higher dimensional ones.
When the resolution is too low to resolve compact dimensions, then the
cross sections  evaluated in the large dimensions are simply multiplied by the volume of the compact
dimensions.} 
The relative probability of events $A$ and $B$ is then given by
\beq
\frac{P_A}{P_B} = \lim_{t\to\infty} \frac{N_A(t)}{N_B(t)},
\label{Psigma}
\eeq
where $N_A(t)$ and $N_B(t)$ are the corresponding numbers of encounters
up to time $t$ along the geodesic.  It does not matter which time variable  is used in Eq.~(\ref{Psigma}), as long as it is monotonic along the geodesic.) 


The cross-section $\sigma_A$ generally depends on the velocity of the watcher relative to the domain of $A$.  So different encounters of the watcher's geodesic with the same type of event will be counted with different weights $\propto \sigma_A^{-1} ({\bf v})$, depending on the velocity ${\bf v}$ of the particular encounter.   The cross-section $\sigma_A ({\bf v})$ is well defined, as long as the domain $D_A$ is sufficiently 
small, so that spacetime curvature on the scale of $D_A$ can be
neglected.  We can then construct a local geodesic congruence parallel to
the watcher's geodesic and define $\sigma_A$ as the volume in the
hyperplane orthogonal to the congruence occupied by the geodesics that
cross $D_A$.\footnote{This prescription can still be applied when
  the curvature gets large in some parts of the domain $D_A$, for
  example, when  
$D_A$ contains compact massive stars or black holes.  All we need to require is that a parallel congruence can be constructed in a small region exterior to $D_A$.  The congruence does not have to remain parallel after crossing $D_A$.} 

In cases when the curvature is not negligible, for example, when the spatial extent of $D_A$ is comparable to the horizon, or when its temporal extent is comparable to the Hubble time, a parallel congruence cannot generally be constructed, and the definition of the cross-section becomes ambiguous. 

To remove this ambiguity we may proceed as follows. Suppose we wish to
define the cross section $\sigma_A$ of a story whose spatial extent is
comparable to the horizon size, with accuracy $(\delta
\sigma/\sigma_A)\sim e^{-3N}$, for some large given positive $N\gg 1$.
Let $t_A$ be the time at which the watcher's geodesic first encounters the story
$A$. Let us now consider an earlier time $t_e< t_A$ along the geodesic
such  that  
\begin{equation}
\int_{t_e}^{t_A} \Theta dt = 3N. \label{te}
\end{equation}
Here $\Theta$ is the expansion of a congruence (of very small width) which is parallel to the geodesic at time $t_e$.
If Eq. (\ref{te}) has more than one solution for $t_e$ (as may happen if there are caustics along the geodesic), we then take the solution which is closest to $t_A$.
The cross section of the story can now be defined from the cross section at time $t_e$, corrected by the expansion to its value at $t=t_A$: 
\begin{equation}
\sigma_A \equiv e^{3N}  \sigma(t_e).
\end{equation}
Here $\sigma(t_e)$ is the volume occupied at time $t_e$ by all the geodesics in the congruence that will go through the story $A$ in the future.
This volume is defined on a flat spacelike hypersurface orthogonal to the watcher's geodesic. Strictly speaking, the hypersurface cannot be  
exactly flat. However, in order to determine $\sigma(t_e)$ we need a congruence whose spatial extent is only of order $e^{-N}$ times the horizon size. Parallel congruences can therefore be defined to the required precision.

In the multiverse context, 
the probability $P_A$ of an event $A$ can be expressed as
\beq
P_A \propto \sum_j X_j N_A^{(j)},
\label{PXF}
\eeq
where $X_j$ is the frequency at which vacuum of type $j$ appears in the
sequence of vacua visited by the watcher, $N_A^{(j)}$ is the average number of events of type $A$ encountered during a visit to vacuum of type $j$, and the summation is over all types of vacua.  The
frequencies $X_j$ depend on the transition rates between different
vacua in the multiverse; we shall set up a formalism for calculating
$X_j$ in the following section.  The quantities $N_A^{(j)}$, on the other
hand, depend only on the physics in vacuum $j$ and on the average time $\tau_j$ spent in this 
vacuum during one visit.   

The measure prescription (\ref{Psigma}), which we shall call "the
watcher measure", has close similarity to Nomura's single observer
measure and to the fat geodesic measure, but there are also some
differences.  We shall comment on these below, in Subsection
  \ref{nomura}.

\subsection{Guth-Vanchurin paradox vs. $Q$-catastrophe.}

When the stories under consideration are significantly extended over a period of time, comparable to the expansion time or larger, all known measures are afflicted by anomalies such as the youngness bias \cite{Tegmark,Guth07}  and the closely related Guth-Vanchurin paradox \cite{guthvanchurin}. As we shall see, the watcher's measure is not immune to such peculiarities
\footnote{We are grateful to Ken Olum for very useful discussions of the issues addressed in this Subsection.}.

Consider a long story which starts with a Hubble patch $A$ at the beginning of slow roll inflation (soon after a bubble forms) and ends with some specific measurement $B$ which is performed after thermalization. If we allow for such long stories to be included in the watcher's tally, it is clear that they are more likely to be tagged near the beginning $A$ of the story than near the end of it. The reason is that cosmic expansion tends to separate the watcher's geodesic from the location where the experiment $B$ 
is performed. On the other hand, the number of instances of $B$ which can follow from a given $A$ is proportional to the volume of the thermalized region which is generated by inflation from the initial Hubble patch. This volume is proportional to  $e^{3N}$, where $N$ is the number of e-foldings of inflation.  
Thus, the probabilities for the outcomes $B$ contain an exponential dependence on $N$. This leads to a phenomenological tension, which is known as the $Q$-catastrophe \cite{Feldstein,qcatastrophe,Graesser}.
The problem arises whenever the value of an observable parameter, such as the amplitude of primordial perturbations $Q$, is correlated with the number of e-foldings of inflation (as happens when perturbations are generated by the slowly rolling inflaton). In this case, the probability distribution for the observable in question will be pushed, by the exponential dependence, to
values which are only marginally consistent with the existence of observers. Typical observers would then measure a very harsh environment, far less comfortable than the one we see 
around us.

The $Q$-catastrophe can be avoided if we adopt the following prescription. First, we drop the long stories from the watcher's tally, and instead we concentrate on the shorter stories $O_i$ representing
the final stages of the long stories. As in the example above, $O_i$ may be a measurement involving some
equipment, whose size and duration are well below the Hubble scale, and may also include some records of the earlier parts of the long story. In general, $O_i$ should include sufficient detail, so that all stories whose probabilities we want to compare can be identified. 
Now, for a  long story to be counted, we may require that the watcher goes through its final stage $O_i$. As mentioned above, because of dilution by the Hubble expansion, 
it is less probable for the watcher's geodesic  to pierce through $O_i$ than it is for it to tag the beginning of the story. The suppression factor is the inverse of the volume expansion factor since 
the beginning of the long story. In the example where the long story starts at the beginning  of slow roll inflation, this compensates for the volume factor $e^{3N}$ corresponding to the size of the thermalized region, and the $Q$-catastrophe is avoided.

It should be noted, however, that this prescription for dealing with long stories leads immediately to the Guth-Vanchurin paradox \cite{guthvanchurin}. Cosmic expansion makes it less likely for the outcome of a story to be tagged by the watcher's geodesic the longer it takes for this outcome to be produced. Now, let us assume that a process initiated at time $t_0$ can have an inmediate outcome $O_A$ or a delayed outcome $O_B$. Following Ref. \cite{guthvanchurin}, we may consider the situation where the alternatives $O_A$ and $O_B$ consist of a subject ingesting a sleeping pill which causes a short or a long sleep, respectively. Which pill the subject is administered is determined at the time $t_0$ by the toss of a fair coin, so that at the time when she falls asleep, the subject should bet that the two alternatives are equally probable. On the other hand, by the time she wakes up, she is more likely to be tagged by the watcher's geodesic in alternative $O_A$ than in alternative $O_B$, so $O_A$ should receive a higher probability than $O_B$.  The same conclusion applies if we use the scale factor, fat geodesic, or causal patch measures. 

Olum has pointed out \cite{ken} that this conclusion violates the principle that probabilities should not be updated in time unless new information is learned. At present, it is unclear to us how to avoid this effect within the context of the watcher's measure (or any other existing measure for that matter). 
On the other hand, it was argued in \cite{guthvanchurin} that the Guth-Vanchurin paradox does not lead to any phenomenological problems, amounting only to a mild youngness bias (which is present in all known global and local cut-off measures anyway).  As of now, the best available option appears to be the prescription of counting a long story only if the watcher's geodesic pierces through its final stage $O_i$.

\subsection{Relation to fat geodesic and Nomura's measures}

\label{nomura}

The fat geodesic measure \cite{Bousso:2008hz} samples a cylindrical spacetime region centered on the watcher's worldline.  The region is chosen so that its orthogonal cross-section has a fixed physical volume $V$.  
Relative probabilities of events are then given by their relative numbers within the sampling region. 
 This prescription is well defined only if $V$ is sufficiently small, so that local curvature is negligible; otherwise, the orthogonal sections of the cylinder are not uniquely defined. Hence, events of extent comparable to the horizon cannot be counted in this measure.
Even if we restrict to sub-horizon events, there is still a problem.
The watcher's worldline will traverse all types of vacua, so in order
for the fat geodesic to be well defined, its thickness $\delta$
should be smaller than the horizon of the highest-energy vacuum, 
$\delta \ll H_{max}^{-1}$, where $H_{max}$ is the largest Hubble
expansion rate in the landscape.    
This means that $\delta$ is very tiny, much smaller than the size of
an atom, and thus much smaller than the extent of any macroscopic
event that we may be interested in.\footnote{Alternatively, we could
  define the sampling region as a set of points within a fixed
  physical  
distance $\delta$ from the watcher worldline.  The distance could be
measured along spacelike geodesics orthogonal to the worldline.
However, the largest spacelike geodesic separation which is possible
in (approximately) de Sitter space of expansion rate $H$ is $\sim \pi
H^{-1}$.  Hence, in order for the thick geodesic to be well defined,
the thickness $\delta$ should be smaller than $\pi H_{max}^{-1}$.} 

One can adopt the attitude that the worldline needs to be thickened only in low-energy regions with relatively small $H$, where observers can exist.  But here one can run into a problem with Boltzmann brains.  
As discussed in Ref.~\cite{DGLNSV}, Boltzmann brains can in principle have a very small size and can occur in high-energy dS vacua.  The corresponding horizon radius $H^{-1}$ can be as small as, say, 1 cm.  If Boltzmann brains are to be included into consideration, the thickness parameter $\delta$ should satisfy $\delta < \pi H_{BB}^{-1}$, where $H_{BB}$ is the largest value of $H$ consistent with the existence of Boltzmann brains.  This implies $\delta \lesssim 1$~cm, which is smaller than the extent of most relevant events.  

Aside from these complications, the fat geodesic and watcher's measures will be approximately equivalent under certain conditions. 
A necessary condition is that the thickness of the fat geodesic should be large compared with the size of the stories under consideration, so that these can be counted in. The thickness should also be smaller than the clustering scale, so that the region which is sampled has a mean density typical of the regions which will be encountered by the watcher's geodesic. 

We now turn to Nomura's measure proposal \cite{Nomura}.  He suggests that the quantum state of our observable region should be compared to that of the horizon region around the watcher. The probability for our region to have certain features is then proportional to the frequency at which these features are encountered by the watcher.  
Clearly, a direct implementation of this proposal would require a
  quantum theory of gravity.  In the meantime, the following
  prescription is suggested \cite{Nomura}.   
Noting that an important feature of our observable region is the
presence  of a physical observer at its center, one may adopt the rule that in order for an observation to be counted, the watcher's geodesic should pass through the head of the relevant observer at an appropriate time.  Nomura argues that the measure is then equivalent to the fat geodesic measure, with the thickness of the geodesic set to be equal to the average size of the observer's head. 
 
It is not clear, however, how this rule should be applied to an
observation like the measurement of the dark energy density by the
High-Z Supernova Search Team.  The measurement involved a number of
people over an extended period, so it is not clear whose head the
geodesic should go through and at what time.  Should we use the size
of the planet instead of the size of the head?  Ref. \cite{Nomura}
  argues that the predictions of the fat geodesic measure are not
very sensitive to the thickness of the geodesic\footnote{This
  argument ignores gravitational clustering. The watcher's geodesic
  behaves like a particle of cold dark matter, and the density of
  objects around it will depend on whether the thickness of the
  geodesic is below or above the clustering scale in a given region of
  spacetime.}.  We note, however, that the quantity of interest could
be correlated with the size of observers (or planets).  One example is
the gravitational constant, which may take different values in
different parts of the multiverse.  
If we use geodesics of variable thickness, adjusted to the size of the
 observer's heads, then we have a size bias which discriminates against
 petit observers.\footnote{In the case of the watcher's measure, this is compensated
 for by the cross section correction in Eq. (\ref{corrected}).} 

Another potential difficulty with this approach is that it is not clear how it should be applied to Boltzmann brains.  A Boltzmann brain is completely delusional; it may think that it is a part of the High-Z Supernova Search Team, while in fact it could be a tiny contraption in a high-energy dS vacuum.   Should we require that the watcher worldline should cross this contraption at the moment when it is having its dreams?  But the imagined location of the Boltzmann brain in space and time is unrelated to its actual location, and in Nomura's approach it is hard to see why it has to be located at the center of the watcher's observable region.




\section{Rate equations}

We shall now set up a formalism for calculating the frequency of visits to
different vacua and the fraction of time the watcher spends in each vacuum. 

\subsection{de Sitter landscape}

We begin by reviewing the case where the landscape includes only
positive-energy vacua.  (Here and in Section 3.3 we closely follow Ref.~\cite{VV}.)

Let us consider a large ensemble of eternal geodesic observers (watchers).  They
evolve independently of one another, yet statistically all of them are
equivalent. For each watcher, we define the proper time $t$, measured
from some arbitrarily chosen point on the geodesic.
The fraction of watchers $f_j(t)$ located in vacuum of
type $j$ at time $t$ obeys the evolution equation
\cite{GV98}\footnote{We shall assume for simplicity that transitions
  between different vacua occur through bubble nucleation.}  
\beq
\frac{df_i}{dt}=\sum_j M_{ij}f_j ,
\label{rateeq}
\eeq
where the summation is over all vacua,
\beq
M_{ij}=\kappa_{ij}-\delta_{ij}\sum_r \kappa_{ri} ,
\label{M}
\eeq
and $\kappa_{ij}$ is the probability per unit time for a watcher who
is currently in vacuum $j$ to find herself in vacuum $i$.  $f_j$ are
assumed to be normalized as
\beq
\sum_j f_j = 1.
\eeq


The transition rate $\kappa_{ij}$ can be expressed as\footnote{Throughout the paper we use Planck units.  Einstein's summation convention is not used: all summations are explicitly indicated.
{For theories where some of the moduli may affect the local value of Newton's constant, our discussion applies provided that we use the Einstein frame (in order to express parameters such as the 
dS temperature or the transition rates)}.}
\beq
\kappa_{ij}=(4\pi/3)H_j^{-3}\Gamma_{ij} ,
\label{kappa}
\eeq
where $H_j=(8\pi\rho_j/3)^{1/2}$ is the de Sitter expansion rate in vacuum $j$,
$\rho_j$ is the corresponding vacuum energy density, and $\Gamma_{ij}$
is the nucleation rate per unit spacetime volume for bubbles of vacuum
$i$ in parent vacuum $j$.  In a semiclassical expansion, this is given by
\beq
\Gamma_{ij}\approx A_{ij}e^{-I_{ij}-S_j} .
\label{Gamma}
\eeq
Here, $I_{ij}$ is the action of the tunneling instanton \cite{CdL} and $A_{ij}$ is a prefactor arising from 
integration of small perturbations around this saddle point. The factor $e^{S_j}$ is shorthand for the semiclassical 
path integral around the Euclidean de Sitter saddle point corresponding to the parent vacuum. This can be interpreted as the exponential of the entropy of vacuum $j$ (with 
loop corrections included). To lowest order, this is given by the Gibbons-Hawking expression
\beq
S_j= \pi/H_j^2 + ...
\label{entropy}
\eeq
where the ellipsis indicates loop corrections.
The instanton
action and the prefactor $A_{ij}$ are symmetric with respect to
interchange of $i$ and $j$ \cite{LeeWeinberg}.  Hence, we can write 
\beq
\kappa_{ij}=\lambda_{ij}H_j^{-3}e^{-S_j}
\label{lambda}
\eeq
with 
\beq
\lambda_{ij}=\lambda_{ji}.
\label{symm}
\eeq
The transition probabilities $\kappa_{ij}$ have the property
\beq
\kappa_{ij}/\kappa_{ji}=(H_i/H_j)^3 \exp(S_i - S_j).
\label{detailedbalance}
\eeq
The relation between this result and detailed balance will be discussed
in Subsection \ref{detbal}. 

The rate equation (\ref{rateeq}) is usually used to describe a congruence of geodesics, emanating orthogonally from some initial spacelike surface.  This description breaks down in regions of structure formation and in AdS bubbles, where the congruence necessarily develops caustics.  We emphasize that here we consider an ensemble of separate geodesics, which are not assumed to form a congruence.

The asymptotic form of the distribution $f_j(t)$ at late times can be expressed as
\beq
f_j(t)=s_j e^{\beta t} ,
\eeq
where $\beta$ is the eigenvalue of $M_{ij}$ having the largest real part (we shall refer to it as the dominant eigenvalue) and $s_j$ is the corresponding eigenvector.  For an irreducible landscape, it can be shown (see Appendix A) that (i) the matrix $M_{ij}$ has a unique eigenvector $f_j^{(0)}$ with zero eigenvalue, 
\beq
\sum_j M_{ij}f_j^{(0)} = 0 ,
\label{Mf0}
\eeq
and that (ii) the real parts of its other eigenvalues are all
negative.  It follows that any solution of the evolution equation
(\ref{rateeq}) with an arbitrary (positive-semidefinite) initial
distribution approaches 
$f_j^{(0)}$ in the asymptotic future,
\beq
f_j(t\to\infty)=f_j^{(0)} .
\eeq

The stationary distribution $f_j^{(0)}$ can be found explicitely
\cite{VV},
\beq
f_j^{(0)}\propto H_j^3 e^{S_j}. 
\label{f0}
\eeq
This can be easily verified by substituting (\ref{f0}) into
(\ref{Mf0}) and making use of (\ref{lambda}) and (\ref{symm}).

\subsection{Detailed balance and ergodicity}\label{detbal} 

Consider a system with a finite number of microstates, and let us denote by $w_{mn}$ the transition rate from a microstate labeled by $n$ to a different microstate 
$m$.  The probability $P_m$  for the system to be in state $m$ obeys the master equation
\begin{equation}
{dP_m \over dt} = \sum_n w_{mn} P_n -\sum_n w_{nm} P_m.
\end{equation}
We are interested in the stationary solution, $dP_m/dt$ = 0.  Then, assuming
detailed balance,
\begin{equation}
w_{mn} = w_{nm}, \label{diba}
\end{equation}
it is clear that $P_m = const.$ is a solution.  This is the microcanonical
ensemble.  Provided that the set of
quantum states is irreducible, the theorem in Appendix A shows that this is the only solution.

Consider now the transition rate $\kappa_{ij}$ from any microstate in a horizon region of vacuum $j$ to 
any microstate in a horizon region of vacuum $i$. This will be given by  
\begin{equation}
\kappa_{ij} = e^{-S_j} \sum_{mn} w_{nm}. \label{macro}
\end{equation}
Here, $n$ and $m$ runs over all microstates in vacua $i$ and $j$ respectively, and $e^{S_j}$ denotes
the number of microstates in a horizon region of vacuum $j$. In deriving (\ref{macro})  we use that the probability of any 
microstate $m$ (conditioned on belonging to vacuum $j$) is given by $e^{-S_j}$, which follows from the microcanonical distribution.
For the reverse transition, and asuming detailed balance, i.e. Eq. (\ref{diba}), we have
\begin{equation}
\kappa_{ji}= e^{-S_i} \sum_{nm} w_{mn} = e^{S_j-S_i} \kappa_{ij}.
\end{equation}
Aside from the factor $(H_i/H_j)^3$, this coincides with Eq.~(\ref{detailedbalance}).
However, the factor $(H_i/H_j)^3$ in Eq.~(\ref{detailedbalance}) appears to indicate a small deviation from detailed balance. 

Even if this deviation is small, it represents a qualitatively significant departure from conventional wisdom. One might think that quantum gravity corrections to the entropy might compensate for the prefactor $H^{-3}$, thus restoring detailed balance. However, as explained around Eq. (\ref{Gamma}), the exponential 
$e^{-S_j}$ featuring in the nucleation rate already contains loop corrections to the entropy. The factor $H^{-3}$ has a geometric origin \cite{LeeWeinberg}: the rate 
$\kappa$ at which we are likely to be hit by a bubble of a new vacuum is proportional to the nucleation rate per unit volume $\Gamma$ times the volume $H^{-3}$ of the accessible horizon region. This seems to indicate that the deviation from detailed balance is for real, with the rate of transitions from the low energy microstate $m$ to the high energy microstate $n$ being larger than the reverse transition.

Violations of detailed balance indicate a preferred time direction for the transitions between given pairs of microstates, signaling the existence of a global arrow of time. This is perhaps not too surprising. In the inflating multiverse, the dynamics of bubble formation is not time symmetric. Bubbles of a lower energy vacuum are expected to nucleate at rest and subsequently expand into the higher energy vacuum, to asymptotically infinite size. But we do not have contracting bubbles of a low energy vacuum shrinking from arbitrarily large size to zero radius, leaving nothing but false vacuum behind them. The absence of contracting bubbles can be thought of as due to initial conditions. The corresponding arrow of time would then be a persistent effect of this initial condition.

Note that the distribution  $f_j^{(0)}$, given in Eq. (\ref{f0}), can be interpreted as the probability for
a randomly picked watcher in the ensemble to be in vacuum $j$.
Alternatively, it can be  interpreted as the fraction of time
spent by each watcher in vacuum $j$. 
Ignoring the prefactor $H_j^3$, 
the distribution $f_j^{(0)}$ is proportional to the
statistical weight of the corresponding vacuum, $e^{S_j}$.  In this sense, the horizon region of the  watcher exhibits an approximately ergodic behavior.  
This can be attributed to the fact that the transition rates (\ref{detailedbalance}) approximately satisfy detailed balance: it is well known that ergodicity can be derived from the detailed balance condition \cite{Kittel,Huang}.

\subsection{Frequency of visits}

We now turn to the calculation of the frequency at which the watcher visits different vacua.
For this purpose, instead of the proper time, we introduce a discrete time variable, $n = 1,2,3,...$, which is incremented by one whenever the watcher jumps to a different vacuum state.  Let $X_j(n)$ be the fraction of watchers in vacuum $j$ at "time" $n$.  $X_j(n)$ is normalized as
\beq
\sum_j X_j(n) = 1
\eeq 
and satisfies the evolution equation
\beq
X_i(n+1)=\sum_j T_{ij} X_j(n) ,
\label{raten}
\eeq
where the transition matrix is given by
\beq
T_{ij}=\frac{\kappa_{ij}}{\kappa_j}
\label{Tkappa}
\eeq
and
\beq
\kappa_j = \sum_r \kappa_{rj} .
\eeq
The diagonal elements of the transition matrix are exactly zero,
\beq
T_{ii}=\kappa_{ii} = 0,
\eeq
since we require each watcher to jump to some other vacuum at every time step.  

For an irreducible landscape that we are considering here, one expects that the evolution equation 
(\ref{raten}) has a stationary solution satisfying
\beq
\sum_j  (T_{ij}-\delta_{ij})X_j = 0.
\label{x}
\eeq
And indeed, rewriting (\ref{x})  as
\beq
\sum_j M_{ij} (X_j/\kappa_j)=0,
\label{Mx}
\eeq
and comparing with eq.~(\ref{Mf0}), we see that the stationary solution of (\ref{Mx}) is 
\beq
X_j \propto \kappa_j f_j^{(0)}.
\eeq
The quantity $X_j$ is proportional to the frequency at which the vacuum $j$ appears in the sequence of vacua visited by the watcher.  The average time spent in this vacuum during one visit is
\beq
\tau_j = \kappa_j^{-1}.
\label{tauj}
\eeq

\subsection{Including AdS vacua} \label{includingads}

Suppose now that along with dS vacua the landscape includes some AdS vacua.  The frequency equation (\ref{raten}) can be straightforwardly generalized to this case,
\beq
X_I(n+1)=\sum_J T_{IJ} X_J(n) .
\label{freq}
\eeq
Here, capital letters in the indices refer to all vacua, both dS and AdS.  When we need to make a distinction between them, we shall use letters from the middle and from the beginning of the Latin alphabet to label dS and AdS vacua, respectively.  Thus, a more detailed form of Eq.~(\ref{freq}) is
\beq
X_i(n+1)=\sum_j T_{ij}X_j(n) + \sum_a T_{ia}X_a(n) , 
\label{freqds}
\eeq
\beq
X_a(n+1)=\sum_j T_{aj}X_j(n) + \sum_b T_{ab}X_b(n) . 
\label{freqds}
\eeq
Here, transition probabilities from dS vacua are given by the same expression as before,
\beq
T_{Ij}=\frac{\kappa_{Ij}}{\kappa_j},
\label{TIJ}
\eeq
with 
\beq
\kappa_j = \sum_I \kappa_{Ij}
\eeq
and $\kappa_{Ij}$ from Eq.~(\ref{kappa}).  At this stage we make no assumptions about the transition probabilities $T_{Ia}$ from AdS vacua, except that all $T_{IJ}$ should satisfy 
\beq
\sum_I T_{IJ} = 1.
\label{unitarity}
\eeq
Unlike the branching ratios from dS vacua, $T_{Ij}$, which can be calculated from the transition rates in the low energy theory, the branching ratios $T_{Ia}$ from AdS vacua depend on dynamics near the bounce, and therefore are UV sensitive.

The asymptotic form of the distribution $X_J(n)$ at late times can be expressed as
\beq
X_J(n) \propto A_J \gamma^n ,
\eeq
where $\gamma$ is the dominant eigenvalue of $T_{IJ}$ (that is, the eigenvalue having the largest real part) and $A_J$ is the corresponding eigenvector.  On physical grounds, we expect the asymptotic distribution to be stationary, which means that the dominant eigenvalue should be $\gamma=1$.  It can be shown that this is indeed the case; see Appendix A.  Hence, the asymptotic distribution can be found by solving the equation
\beq
\sum_J T_{IJ}X_J=X_I.
\label{stationary}
\eeq

The fraction of time $f_J$ spent by the watcher in vacuum of type $J$ can be expressed as
\beq
{f}_J = \frac {X_J \tau_J}{\sum_I X_I \tau_I},
\label{fJ}
\eeq
where $X_J$ is found from Eq.~(\ref{stationary}) and $\tau_J$ is the average time spent in vacuum $J$, which is given by (\ref{tauj}) for dS vacua and is determined by the classical evolution up to the crunch for AdS vacua.

\subsection{A special case}

An interesting special case is when the transition probabilities $T_{ja}$ from AdS crunches to dS vacua are independent of the crunching vacuum $a$,
\beq
T_{ja}\equiv Q_j .
\eeq
This may be a reasonable assumption: in the extreme conditions of the crunch the nature of the original vacuum may be forgotten.  We shall also assume for simplicity that transitions between AdS vacua do not occur, $T_{ab}=0$.  Then Eq.~(\ref{stationary}) can be rewritten as
\beq
\sum_j (T_{ij}-\delta_{ij})X_j = -\xi Q_i ,
\eeq 
where $\xi=\sum_a X_a$, or in the matrix form
\beq
({\tilde T}-I){\tilde X}=-\xi Q.
\label{TXP}
\eeq
Here, ${\tilde X}$ and $Q$ are $N$-vectors and ${\tilde T}$ is an $N\times N$ matrix, where $N$ is the number of dS vacua, and $I$ is the unit matrix in the same vector space.  The vector ${\tilde X}$ includes only dS components $X_j$ of the distribution $X_J$, and similarly ${\tilde T}$ includes only the matrix elements $T_{ij}$ between dS vacua.  The solution of Eq.~(\ref{TXP}) for ${\tilde X}$ is
\beq
{\tilde X}=\xi (I-{\tilde T})^{-1}Q.
\label{XTP}
\eeq
The constant $\xi$ can be determined from the normalization condition,
\beq
1=\sum_J X_J = \sum_j X_j +\xi,
\eeq
with $X_j$ from (\ref{XTP}).  Once $X_j$ are found, the frequencies of visits to AdS vacua can be determined from
\beq
X_a = \sum_j T_{aj}X_j.
\label{Xa}
\eeq

We note that a solution of the same form (\ref{XTP}) was obtained by Vanchurin \cite{Vanchurin} in a different context.  He assumed that AdS vacua are terminal and considered an ensemble of (non-eternal) observers with an initial distribution $P_j$.  $X_j$ is then defined as the number of times the vacuum $j$ is visited by all observers in the ensemble.  The relation between this setup and ours  is  not difficult to understand.  In Vanchurin's context, instead of starting the observers' histories at the same `time' $n=0$, we can follow them sequentially.  For an infinite ensemble, the resulting history can be thought of as a history of an eternal observer.  Specifically, the construction can be pictured as follows.  First we draw an observer from the initial distribution $P_j$.  We follow his evolution until he hits the crunch in some AdS vacuum.  We then draw another observer from $P_j$ and attach his history as a continuation of the first observer's history.  After the second observer hits the crunch, we return to the initial distribution again, and so on.   In our picture, every time an eternal observer gets into an AdS vacuum, he continues after the bounce in a dS vacuum $j$ with probability $Q_j$.  Clearly, this should give the same frequencies $X_j$ if we identify $Q_j$ with Vanchurin's initial distribution $P_j$.

\subsection{A mini-landscape}

\begin{figure}
\begin{center}
\vspace{-2cm}
\includegraphics[width=14cm]{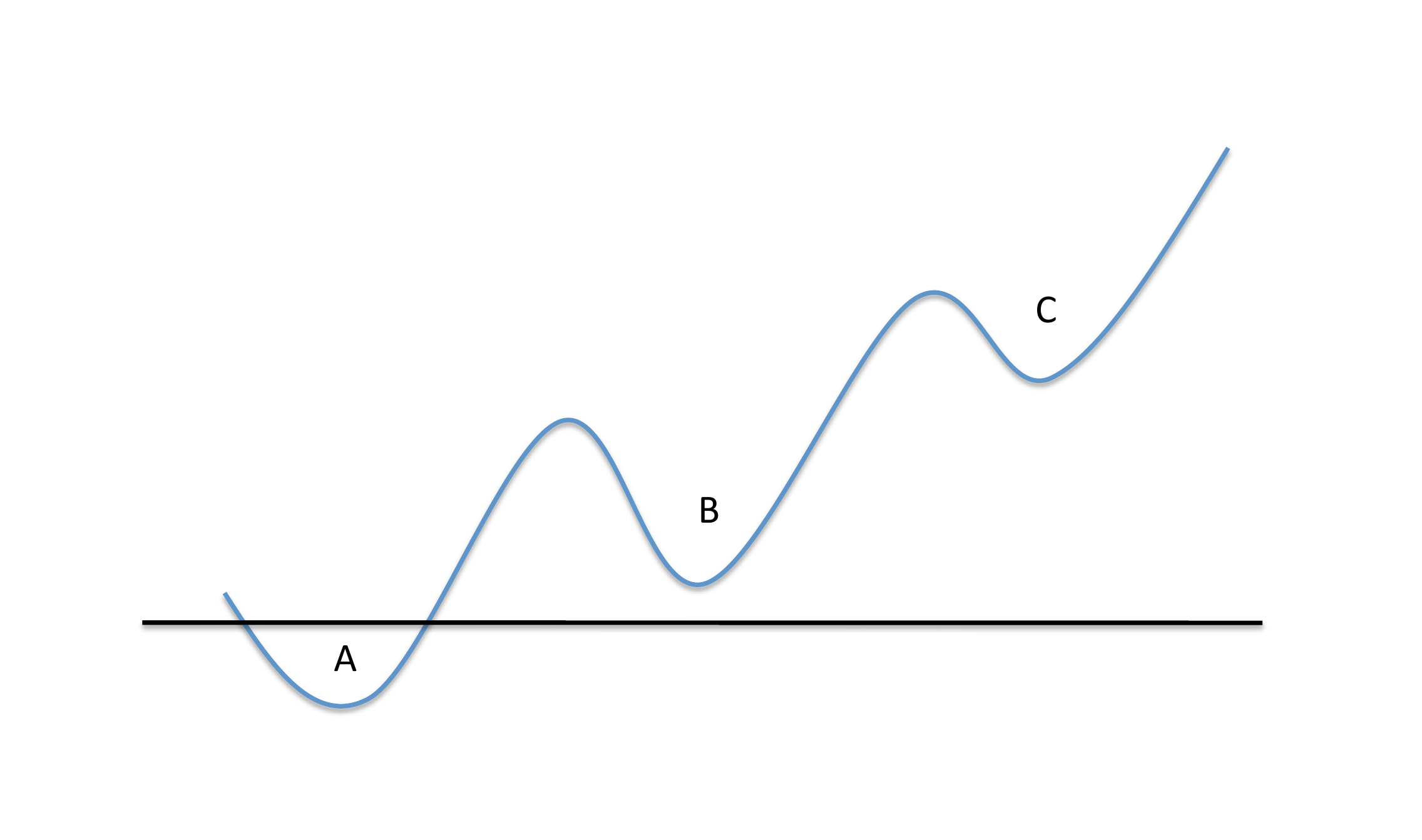}
\vspace{-1.5cm}
\caption{Mini landscape with an AdS vacuum A, a low-energy dS vacuum B, and a high-energy dS vacuum C.  \label{mini}}
\end{center}
\end{figure}

To illustrate the effect of AdS bounces on the probability distribution, we shall consider a simple landscape consisting of just three vacua: an AdS vacuum A, a low-energy dS vacuum B, and a high-energy dS vacuum C (see Fig.~\ref{mini}).  Possible tunneling transitions in this landscape are described by the `schematic'
\beq 
A \leftarrow B \leftrightarrow C .
\label{schematic}
\eeq
AdS crunches in vacuum A are followed by bounces with transitions to B or C.  We shall denote the corresponding probabilities by $Q_B$ and $Q_C$, respectively, with $Q_B + Q_C = 1$.  

The rate equations (\ref{stationary}) for the frequency of visits in this landscape are
\beq
X_A=T_{AB}X_B,
\eeq
\beq
X_B=T_{BC}X_C+Q_B X_A,
\eeq
\beq
X_C=T_{CB}X_B + Q_C X_A,
\eeq
and we find
\beq
X_C/X_B = T_{CB}+Q_CT_{AB}, 
\eeq
\beq
X_A/X_B = T_{AB}.
\eeq
The fraction of time spent by the watcher in different vacua can now be found from Eq.~(\ref{fJ}), $f_J\propto X_J \tau_J$.

The transition probabilities $T_{ij}$ can be expressed in terms of the rates $\kappa_{ij}$ from the definition (\ref{Tkappa}),
\beq
T_{AB}=\frac{\kappa_{AB}}{\kappa_{AB}+\kappa_{CB}}, ~~~ T_{CB}=\frac{\kappa_{CB}} {\kappa_{AB}+\kappa_{CB}},
\eeq
and the average times spent during one visit are given by 
\beq
\tau_B=\frac{1}{\kappa_{AB}+\kappa_{CB}}, ~~~ \tau_C=\frac{1}{\kappa_{BC}},
\eeq
while $\tau_A$ is determined by the classical AdS evolution.  Combining all this, we obtain
\beq
\frac{f_C}{f_B}=\frac{\kappa_{CB}}{\kappa_{BC}}+Q_C\frac{\kappa_{AB}}{\kappa_{BC}},
\label{C/B}
\eeq
\beq
\frac{f_A}{f_B}=\tau_A \kappa_{AB}.
\eeq
Note that the second term in (\ref{C/B}) is important when $Q_C \kappa_{AB}\gtrsim \kappa_{CB}$, that is,  when the rate of transitions from B to C through a bounce at A is comparable to or higher than the rate of direct upward transitions.  In the limit when bounce transitions to C are highly unlikely, $Q_C\to 0$, Eq.~(\ref{C/B}) gives a thermal distribution, $f_C/f_B=\kappa_{CB}/\kappa_{BC}\sim\exp(S_C-S_B)$.  On the other hand, if $Q_C\sim 1$ and B has a much lower energy density than C, so that $S_B\gg S_C$, then the first term in (\ref{C/B}) is negligible and
$f_C/f_B\sim \kappa_{AB}/\kappa_{BC}$.  In this case, the ratio $f_C/f_B$ is not suppressed by the small upward transition rate and can be much greater than $\exp(S_C-S_B)$.

In the absence of AdS bounces, the mini-landscape (\ref{schematic}) was discussed in Ref.~\cite{DGLNSV}.  The outcome then depends on the relative lifetime of the vacua B and C.  For $\tau_B \ll \tau_C$, one finds $f_C/f_B\approx \tau_C/\tau_B$.  In the opposite (and apparently more realistic) case, when the high-energy vacuum has a shorter lifetime, $\tau_C\ll\tau_B$, the result is $f_C/f_B \sim \exp(S_C-S_B)$.

\section{Thermal death vs. the arrow of time}
 
The observed arrow of time presents a potential danger for the eternal geodesic picture.  This is particularly evident in a dS landscape with no AdS vacua.   It has often been argued that a causal patch in such a landscape evolves like a closed Hamiltonian system having maximum entropy $S_{max} =\pi/H_{min}^2$, where $H_{min}$ is the Hubble rate in the lowest-energy dS vacuum.  This view was first proposed by Dyson, Kleban and Susskind \cite{Dyson} and was adopted in much of the subsequent work. It suggests that the evolution of the causal patch is ergodic, so at late times it should be in the state of thermal equilibrium, described by the microcanonical ensemble.  This has a rather unsettling consequence, that the observed state of the universe is most likely to arise as a quantum fluctuation in dS space \cite{Dyson}.  Banks has argued that similar conclusions should apply even in the presence of AdS vacua \cite{Banks}.

Here, we do not adopt the picture of a causal patch as a closed system.    Information continuously escapes the causal patch through the horizon; this information needs to be traced over, resulting in a stochastically evolving density matrix \cite{BoussoSusskind}.  This, however, does not necessarily help to avoid the thermal death problem.   For a dS landscape, the distribution (\ref{f0}) that we obtained from the rate equation is essentially the microcanonical distribution.  It assigns the highest probability to the lowest-energy dS vacuum.  The most likely way to get from that vacuum to our present state is through a `thermal' dS fluctuation.  This leads immediately to the problem of BB dominance. The probability of forming normal observers by tunneling up to a high-energy vacuum with subsequent inflation and standard hot big bang evolution is negligibly small by comparison. (This will be discussed in detail in a forthcoming paper \cite{GGV12}.)

As we mentioned in the preceding section, the approximate microcanonical nature of the distribution (\ref{f0}) can be traced to the approximate detailed balance property (\ref{detailedbalance}) of the Coleman-DeLuccia transitions between dS vacua.  However, there seems to be no reason to expect that bounce transitions after AdS crunches should satisfy detailed balance, not even approximately.  To explore the range of possible qualitative behaviors, let us consider the special case when the transition rates after the crunch are independent of the parent AdS vacuum, $T_{ja}=Q_j$.  In this case we found in Sec.~3 that the evolution of the watcher can be pictured as a sequence of histories with initial data drawn from the distribution $Q_j$.  

Suppose first that the distribution $Q_j$ has the form
\beq
Q_j \propto \exp(-S_j),
\label{tunneling}
\eeq
similar to that obtained from the tunneling \cite{tunneling,valery,alexei} or Linde's \cite{tunneling2} wave function of the universe.  This favors low-entropy states after the crunch.  The watcher then typically observes a succession of dS vacua with lower and lower energies (and higher entropies), ending with a crunch, followed by another dS sequence, etc.  Some of these dS episodes may include inflation, structure formation, and evolution of observers. 

Alternatively, consider a distribution of the Hartle-Hawking form \cite{HH},
\beq
Q_j\propto \exp(+S_j).
\label{HH}
\eeq
In this case, the starting states after the AdS crunches will tend to be the lowest-energy (highest-entropy) dS vacua. In order to have inflation, resulting in a region like ours, the subsequent history  
must include an upward transition to a high-energy vacuum.  Such transitions are double-exponentially suppressed, so in this scenario observers are much more likely to appear as quantum vacuum fluctuations.

\begin{figure}
\begin{center}
\vspace{-2cm}
\includegraphics[width=14cm]{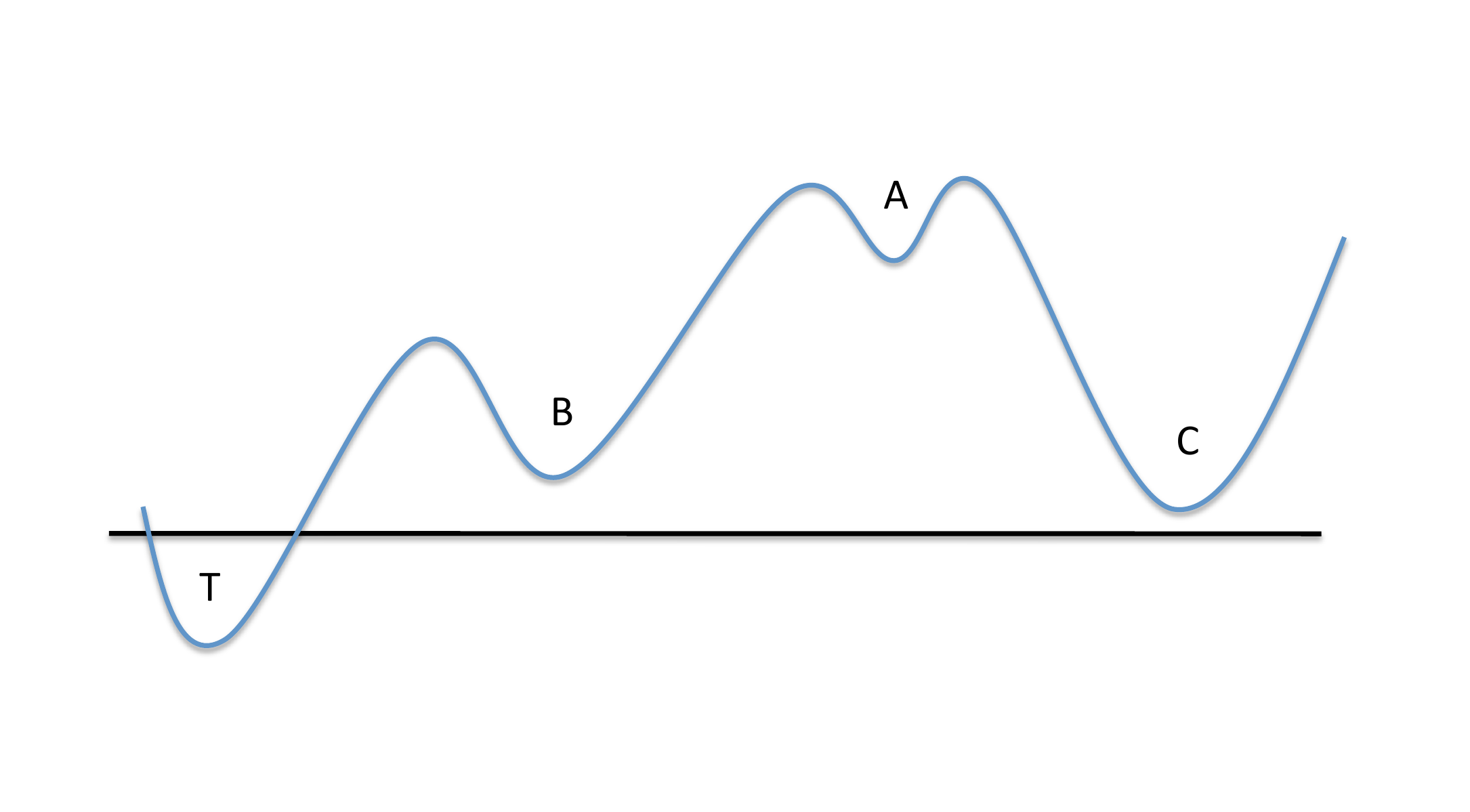}
\vspace{-1.5cm}
\caption{Mini landscape where the anthropic vacuum $B$ is separated from the lowest-energy dS vacuum $C$ by a high-energy metastable  
vacuum state $A$. \label{mini2}}
\end{center}
\end{figure}

Bousso \cite{Bousso12} has pointed out that in some special
landscapes, ordinary observers like us may evolve with a high
probability, even with a Hartle-Hawking (HH) initial
conditions\footnote{Bousso's scenario in \cite{Bousso12} is different
  from what we are discussing here in that he considered observers
  with an initial HH distribution and treated AdS vacua as terminal.
  However, as we discussed in Sec.~2, this is equivalent to our model
  with $T_{ja}=Q_j$. We are grateful to Raphael Bousso for a
    clarifying discussion of this model.}. Consider for instance the mini-landscape
illustrated in Fig.~\ref{mini2}. 
Here, the anthropic vacuum $B$ is separated from the lowest-energy dS
vacuum $C$ by a high-energy metastable   
vacuum state $A$ , and it is assumed that the tunneling transitions
between vacua can occur only to nearest neighbors.  With a HH
distribution, the most likely initial state is in vacuum $C$, 
which is assumed to be unsuitable for life (even in the form of
Boltzmann brains).  The only way to get from $C$ to the anthropic
vacuum $B$ is by first making the highly suppressed transition  to
$A$.  The subsequent transition from $A$ to $B$ 
is accompanied by inflation and by evolution of ordinary observers, with
$B$ eventually decaying to the AdS vacuum $T$.
A special feature of this model is that all histories starting in the
lowest-energy vacuum $C$ encounter ordinary observers with unit
probability. The reason is that the branching 
ratio $T_{AC}=\kappa_{AC}/\kappa_C$ is equal to $1$ in spite of the
smallness of $\kappa_{AC}$. 
In a more general case, when $C$ is allowed to have other decay channels, the
situation is likely to be very different. For instance, if vacuum $C$
is allowed to have a channel of decay into an AdS vacuum $U$, then the
branching ratio $T_{AC}$ is likely to be be very small, due to the
smallness of $\kappa_{AC}\sim e^{-I_{CA}-S_C}=e^{S_A-S_C}
\kappa_{CA}$, compared to $\kappa_C \sim \kappa_{UC}$ (note that both
$\kappa_{CA}$ and $\kappa_{UC}$ correspond to downward transitions,
and do not contain any huge entropy suppression factors).  
Denoting by $\Gamma_{BB}$ the rate of Boltzmann brain (BB) production
in vacuum $B$, we expect this landscape to be dominated by BB provided
that  
$e^{S_B} \Gamma_{BB}/\kappa_B  \gg e^{S_C} T_{AC} \sim e^{S_A}
(\kappa_{CA}/\kappa_{UC})$, which can easily be satisfied due to the
smallness of the entropy in vacuum $A$ compared to the entropy in
vacuum $B$. 

We conclude that, for a generic landscape, transition probability distributions like (\ref{HH}), favoring high-entropy states, contradict observations and should be ruled out.  `Tunneling-type' distributions (\ref{tunneling}), which favor low entropies, are allowed, but this choice is in no way unique.   For example, `entropy-neutral' distributions like $Q_j=const$ are also phenomenologically acceptable.

\section{Extensions}

\subsection{Minkowski vacua}

\begin{figure}
\begin{center}
\vspace{-2cm}
\includegraphics[width=14cm]{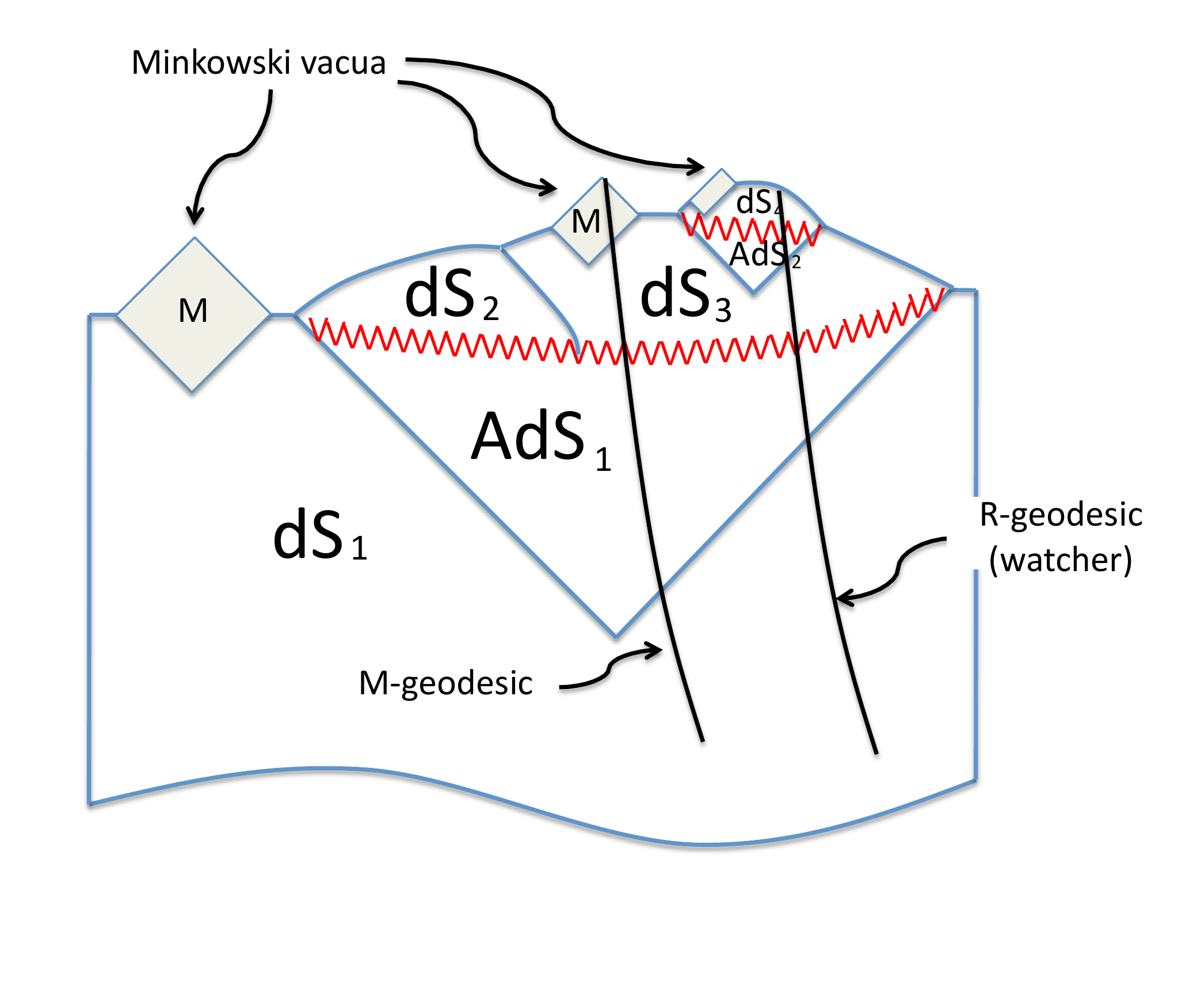}
\vspace{-1.5cm}
\caption{Causal diagram for a multiverse with AdS bounces and terminal Minkowski vacua. \label{terminal}}
\end{center}
\end{figure}

Suppose now that the landscape includes some stable Minkowski vacua (we shall call them $M$-vacua), as suggested by string theory.  Then all timelike geodesics, except a set of measure zero, will end up in these terminal vacua, with their endpoints at Minkowski timelike infinity (see Fig. \ref{terminal}).  We shall refer to them as $M$-geodesics. The remaining, measure-zero geodesics which are constantly recycling between different vacua will be called $R$-geodesics.  The set of vacua visited by any $M$-geodesic is finite and depends on where the geodesic started.  Hence, $M$-geodesics are not useful for defining probabilities (unless we re-introduce an ensemble with some distribution of initial data).

The strategy that suggests itself in this case is to determine probabilities by following one of $R$-geodesics.\footnote{As already mentioned, our approach is in some sense opposite to that adopted in the census taker measure, which is focused entirely on $M$-geodesics.}
It does not matter which geodesic we choose.  With the assumption (to be refined below) that the geodesic does not visit any $M$-vacua, the probabilities $P_J$ and the frequencies $X_J$ can then be found from Eqs.~(\ref{PXF}) and (\ref{stationary}), respectively, with the transition probabilities $T_{IJ}$ given by (\ref{TIJ}) and the indices $I,J$ running only through non-terminal vacua.  Note that Eq.~(\ref{fJ}) for the fraction of time spent by the watcher in different vacua can no longer be used here.  The reason is that the lifetimes of dS and AdS vacua are influenced by their decays into $M$-vacua, which are ignored in this case.

With this prescription, vast Minkowski regions of spacetime are excluded from consideration, and any phenomena that might occur in such regions are assigned zero probability.  This can be justified by observing \cite{SusskindBook} that supersymmetric $M$-vacua cannot support nontrivial chemistry, and thus no measurements can take place in such vacua.  On the other hand, early evolution of $M$-bubble interiors may include periods of inflation and matter domination, during which supersymmetry will be broken.   Observers could exist during such periods, so it does not seem right to exclude them.

In fact, the potentially habitable regions of $M$-bubbles are {\it not} excluded when we restrict to $R$-geodesics, as suggested above.  In such broken-supersymmetry regions, bubbles of other vacua can form and expand.   
So a geodesic entering an $M$-bubble from some parent vacuum can explore part of the habitable region and then exit to some other dS or AdS bubble.  We do not attempt to modify the rate equation to incorporate this effect in the present paper, although such modification should not be difficult.  We note also that similar modifications will generally be required even in purely dS or dS - AdS landscapes.  The transition rates $\kappa_{ij}$ are assumed to be constant in Eqs.~(\ref{rateeq}), (\ref{M}), but they are generally time-dependent during the early stages of bubble evolution.  For a dS vacuum, the rate gradually approaches its constant asymptotic value.  The main difference in the case of $M$-vacua is that the asymptotic rate is equal to zero.

\begin{figure}
\begin{center}
\vspace{-2cm}
\includegraphics[width=14cm]{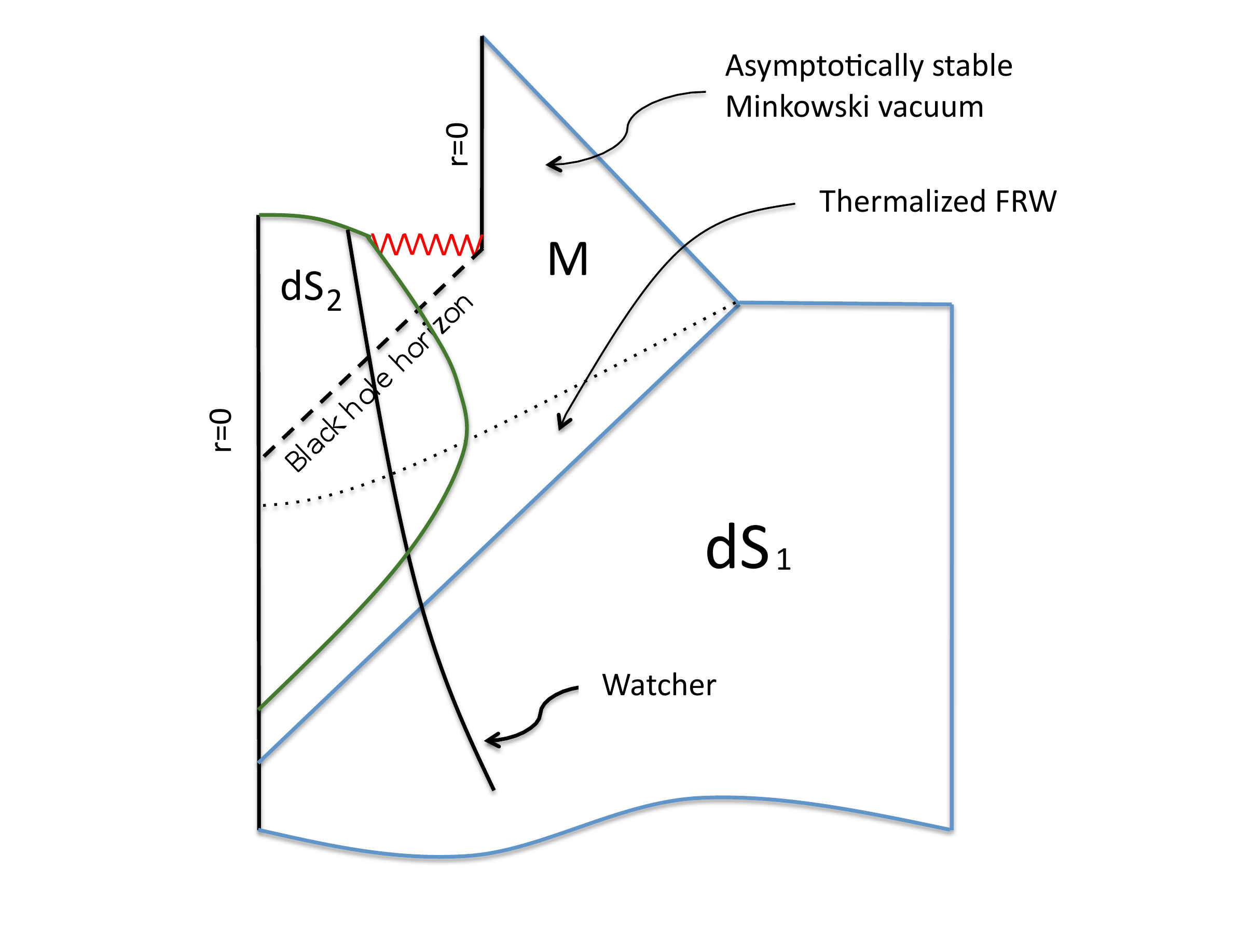}
\vspace{-1.5cm}
\caption{A bubble of positive energy density vacuum, labeled as dS$_2$ in the diagram, may nucleate during a period of slow roll  inflation, or during the thermalized phase that follows immediately after an M bubble nucleates. Initially, the daughter dS$_2$ bubble will expand, just as it would in a parent dS vacuum. It will continue to expand even after the background energy density drops below that in the daughter bubble -- as long as the bubble radius remains larger than the local horizon. The region surrounding the daughter bubble cools down and curvature starts dominating. Eventually, the daughter bubble falls within the horizon and starts contracting, which leads to the formation of a black hole. Finally, the black hole evaporates, leaving behind two disconnected regions of space, corresponding to dS$_2$ and asymptotically stable $M$ vacuum.
{Actually, as we shall discuss in Section \ref{bhsec}, if AdS singularities are resolved in the fundamental theory, the same is likely to apply to black hole singularities. If so, then the wavy line in the above Figure is
not part of the future boundary of the causal diagram, but a gateway leading to new regions of spacetime occupied by other vacua}.}
 \label{contracting}
\end{center}
\end{figure}

The spacetime structure of bubbles formed inside a parent $M$-bubble is somewhat unusual.  Suppose, for example, that there is an early period of inflation in the $M$-bubble and that a daughter dS bubble has nucleated during this period.  Initially, the daughter bubble will expand, just as it would in a parent dS vacuum.  It will continue to expand even after the background energy density drops below that in the daughter bubble -- as long as the bubble radius remains larger than the local horizon.  However, the bubble radius asymptotically grows as $R\propto a(\tau) \approx \tau$, and the horizon radius is
\beq
h(\tau) = a(\tau) \int_0^\tau \frac{d\tau'}{a(\tau')} \approx \tau \ln \tau ,
\eeq
where $a$ is the scale factor and $\tau$ is the FRW time in the parent bubble.  The ratio of the two radii is $h/R (\tau\to\infty) \sim \ln \tau$, and thus the bubble radius inevitably becomes smaller than the horizon.  At that point the daughter bubble begins to contract and eventually collapses to a black hole.  
In the meantime its interior continues to expand and to form its own daughter bubbles.  After the black hole eventually evaporates, this interior becomes a separate inflating multiverse.  A spacetime diagram illustrating this situation is shown in Fig. \ref{contracting}.

.

We thus see that the multiverse generally has a rather complicated spacetime structure, and includes a multitude of spatially disconnected regions.  
\footnote{This spacetime structure is similar to that discussed by Sato, Kodama, Sasaki and Maeda \cite{SKSM} in a somewhat different context.}
The watcher's geodesic can transit from one such region to another through daughter bubbles nucleating inside $M$-bubbles. Each geodesic will generally visit an infinite number of disconnected regions.\footnote{Transitions  to spatially disconnected regions can also occur through spontaneous nucleation of black holes \cite{BoussoBH}; see below.}

\subsection{Reducible landscapes}

Throughout the paper we assumed that the landscape of vacua is irreducible, so that any vacuum is accessible through bubble nucleations and/or AdS crunches from any other vacuum.  If instead 
the landscape splits into several disconnected sectors, the watcher measure can be used to determine the probability distribution $P_J^{(A)}$ in each of the sectors (labeled by $A$).  The full distribution is then given by
\beq 
P_J = Q_A P_J^{(A)},
\eeq
where $Q_A$ is the probability of being in sector $A$.  Since different sectors are inaccessible from one another, the distribution $Q_A$ should depend on the initial state of the universe.  It is natural to expect that it is determined by the wave function of the universe.

\subsection{Terminal AdS vacua}

We finally consider the possibility that AdS vacua are in fact terminal, ending  in true singularities, or, at the quantum level, in domains where classical space and time cease to exist. 
It appears that the only way to apply the watcher measure to a landscape with terminal vacua is to use an ensemble of geodesics.  One then has to address the problem of choosing the initial distribution $Q_J$ for this ensemble.  Once this distribution is determined, the frequencies $X_J$ can be found from Eqs.~(\ref{XTP}),(\ref{Xa}).

Once again, a natural choice might be to derive the initial distribution from the wave function of the universe, $\Psi$ \cite{GV98}. This wave function unfortunately is itself rather uncertain, and one can get vastly different distributions, depending on one's choice of the boundary conditions for $\Psi$.  The two most popular choices are the tunneling \cite{tunneling,tunneling2} and Hartle-Hawking \cite{HH} wave functions, which give the distributions (\ref{tunneling}) and (\ref{HH}), respectively.  As we discussed in the preceding section, the Hartle-Hawking distribution (\ref{HH}) is phenomenologically unacceptable in the present context, while the tunneling distribution (\ref{tunneling}) appears to work fine.

\section{Black holes}

\label{bhsec}

If AdS singularities are resolved in the future fundamental theory, the same is likely to apply to black hole singularities.  This may have important implications for the measure problem, as we shall now discuss.

Black holes can spontaneously nucleate in de Sitter space \cite{Ginsparg,Mellor,Mann,BoussoBH,BoussoHawkingBH}, at a certain rate per unit spacetime volume, and thus the watcher's geodesic has some probability to encounter a black hole per unit proper time.  
As it enters a black hole, the geodesic hits the high-curvature region replacing the singularity and transits to another dS or AdS vacuum.  The possibility of such transitions has been discussed, e.g., in Refs.~\cite{Frolov,Barrabes} in the context of the maximal curvature hypothesis.  The spacetime structure of the multiverse in this case is similar to that discussed in Sec.~5.1. 


\subsection{Transition rate}

At the formal level, watcher's transitions between different vacua through black holes are not much different from transitions through bubble nucleation and can be accounted for by a slight extension of the formalism of Sec.~3.  For example, 
the rate equation (\ref{rateeq}) remains unchanged, while the transition rates are now given by\footnote{Black holes can also be formed by gravitational collapse in structure formation regions.  Accounting for such black holes will require an extension of the formalism similar to that which is necessary to account for a time-varying rate of daughter bubble nucleation at early FRW times inside a parent bubble.}
\beq
\kappa_{ij}=\kappa_{ij}^{bubble}+\kappa_{ij}^{BH},
\eeq
where $\kappa_{ij}^{bubble}$ is the transition rate through bubbles, given by Eqs.~(\ref{kappa}), (\ref{Gamma}), and $\kappa_{ij}^{BH}$ is the transition rate through black holes.  

To estimate the latter, we first note that for sufficiently small black holes the nucleation rate per unit spacetime volume is given by \cite{Mellor,Mann,BoussoHawkingBH}
\beq
\Gamma_{BH}(M) \propto e^{-M/T_{GH}} ,
\label{GammaBH}
\eeq
where $M$ is the black hole mass and $T_{GH} = H/2\pi$ is the Gibbons-Hawking temperature of de Sitter space.  This applies when $M\ll H^{-1}$, that is, when the Schwarzschild radius is small compared to the dS horizon.  The probability for the watcher to be captured by a black hole of mass $M$ per unit time is 
\beq
\kappa_{BH}(M) \sim r_*^3(M) \Gamma_{BH}(M),
\label{capture}
\eeq
where $r_*(M)$ is the maximal distance from the watcher at which a black hole can nucleate and still capture the watcher's geodesic.  The full transition rate $\kappa_{ij}^{BH}$ can be obtained by multiplying the capture rate in Eq.~(\ref{capture}) by the transit probability $T_{ij}$ through the high-curvature region.

The capture radius $r_*$ is given by
\beq
r_*(M) \sim M^{1/3} H^{-2/3},  
\label{r*}
\eeq
provided that the time $t_* \sim H^{-1}$ that it takes for the watcher to fall from $r\sim r_*$ to the black hole is shorter than the black hole evaporation time, $\tau_M\sim M^3$.  This gives the condition for the mass $M > H^{-1/3}$.  For smaller black holes, $r_*(M) \sim M^{7/3}$.  The smallest black holes that we can meaningfully talk about are Planck-scale black holes with $M\sim 1$, for which the lifetime is comparable to the light crossing time.  

Eq.~(\ref{r*}) for the capture radius assumes also that the black hole nucleates at rest relative to the watcher, while in fact it is expected to have a thermal velocity, $v \sim (T_{GH}/M)^{1/2}$.  Then, instead of hitting the black hole head on, the geodesic could orbit around it and avoid capture.  This effect can be significant for $r_* \gtrsim M^{3/2}H^{-1/2}$.  We note, however, that details of the function $r_*(M)$ are not very important, considering that the $M$-dependence of the capture probability (\ref{capture}) is dominated by the exponential factor in $\Gamma_{BH}$.

\subsection{UV cutoff sensitivity}

As mentioned in Subsection \ref{includingads}, the branching ratios for transitions $T_{Ia}$ from AdS vacua to all other vacua depend on UV physics. The same will be true for bounces at black hole singularities, which can also make the branching ratios $T_{Ij}$ from dS vacua to other vacua depend on high energy physics. In fact, the situation is more interesting in the case of black holes, because the decay rates of dS vacua (and not just the branching ratios) can be UV sensitive. 

Indeed, it follows from (\ref{GammaBH}), (\ref{capture}) that the highest capture rate is obtained for the smallest black holes.  This has the consequence that these rates have an exponential dependence on the high-energy cutoff of the theory.
The tunneling suppression factor for black hole nucleation is given by (\ref{GammaBH}), while the corresponding factors for bubble nucleation range from $\sim 1$ to $\sim \exp(-\pi/H^2)$ and can be much smaller than (\ref{GammaBH}) if $M \ll H^{-1}$.  This means that transitions between vacua through microscopic black holes can be much more frequent  than transitions through bubbles, at least for some pairs of vacua.  Indeed, suppose we introduce a short distance cutoff $\xi$.  Then black holes of mass $M\lesssim \xi$ should be excluded from consideration, and the black hole nucleation rate is $\kappa_{ij}^{BH} \sim \exp(-\xi/T_{GH})$, which is exponentially sensitive to the cutoff $\xi$.  We shall now indicate some possible ways of dealing with this unusual situation.

{\it (i) Exclude transitions through black holes.} One possible attitude might be to require that the watcher's geodesic should remain in the same connected component of the multiverse.  Geodesics captured by black holes end up, after evaporation, in regions of space which are disconnected from the asymptotic region in which  the black hole formed.  So, the prescription could be that a geodesic that was caught by a black hole is continued from the point of evaporation after the black hole disappears.  

However,  if we introduce this rule, then we cannot handle landscapes which include stable Minkowski vacua (at least, in the way which we described in Section 5.1). The reason is that dS bubbles which nucleate inside of M-bubbles end up inside of black holes, as illustrated in Fig. \ref{contracting}.  In fact, this prescription seems hard to implement even if we try to enforce it. This can be seen, again, in the example illustrated in Fig. \ref{contracting}. 
The asymptotic dS$_2$ region is causally disconnected from the endpoint of black hole evaporation. So it is unclear 
at what time should the geodesic inside of the dS$_2$ region be discontinued, and reattached to the endpoint of evaporation of the black hole. Note also that this difficulty would arise more generally than just for the case of dS bubbles nucleating in M-vacua. Some high energy dS bubbles can nucleate during the slow roll inflationary phase (or during the subsequent thermalized phase) at the early stages of evolution inside of dS bubbles of a much lower vacuum energy. As the energy of the environment gradually decreases, some of these high energy dS bubbles will collapse to black holes, since their size can be much smaller than the size of the horizon in the low energy dS vacuum. 

Finally, one may argue that a distinction could be made between two different cases. If the watcher falls into
the singularity of the black hole, we could then adopt the rule that at that time its geodesic should be continued at
the endpoint of black hole evaporation, 
while if the watcher escapes into an inflating region that pinches off, as in Fig. \ref{contracting}, then her worldline continues unimpeded and no reattachment is necessary. Nonetheless, if we are accepting that geodesics can pass through singularities, this distinction seems quite artificial.

{\it (ii) Relegate to Quantum Gravity.}  Some approaches to quantum gravity, in particular the holographic ideas, suggest that in quantum theory the multiverse should be described in terms of the wave function of a region encompassed by an apparent horizon surface (e.g., \cite{census,Nomura, BoussoSusskind}).  
In this approach, the geodesics representing the possible trajectories of a watcher may not play any
fundamental role (except perhaps in some appropriate limit), while the apparent horizon would be a more relevant object to consider. Now suppose a small black hole nucleates within a dS vacuum, and then evaporates. If the black hole is small compared to the dS horizon, the cosmological apparent
horizon remains practically unchanged throughout this process. In this picture, it seems plausible to conclude that the feature of UV-sensitivity due to the relatively large nucleation rate of mini-black holes may be irrelevant. The mini black holes may play the role of transient fluctuations in the larger system, but may be completely unrelated to transitions to 
other vacua. In the case of large black holes, or in the case of bubble nucleation, 
the picture may get more complicated and the concept of a watcher may perhaps arise as an effective one.
Implementation of this approach, however, would require a better understanding of quantum gravity.

{\it (iii) Allow transitions through black holes.}  If the watcher's geodesic is allowed to go through mini-black holes, the  decay rates of dS vacua become highly sensitive to UV physics.  
Then we cannot impose a floating high-energy cutoff (as it is usually done in renormalization group applications).  Instead, we should use the true physical cutoff of the theory, e.g., the 
Planck scale. In this case, most black hole transitions will go through Planck-size black holes.  

In the first two approaches, black hole nucleation has no effect on the measure and can be ignored.  In the third approach, it has a rather strong quantitative effect: transition rates between the vacua get significantly modified, resulting in significant changes in the probabilities.  

At the qualitative level, 
the third approach may help to resolve the Boltzmann brain problem of eternal inflation.   The probability of nucleating a Boltzmann brain of mass $M$ is \cite{Page07,DGLNSV,Bousso:2008hz} 
$\sim \exp(-M/T_{GH})$, and for $M$ large compared to the Planck mass, it is much smaller than the probability of forming a Planck-size black hole.  Hence, the watcher's geodesic is likely to encounter a black hole and exit to a disconnected component of the multiverse before it encounters a Boltzmann brain.  On the other hand, when the geodesic passes through habitable parts of the multiverse, it is likely to encounter ordinary observers who evolved by natural selection, provided that the time $\tau_{obs}$ it takes to 
evolve observers is less than the typical time $\tau_{BH}$ that it takes to encounter a black hole.
For Planck-size black holes, $M\sim 1$, the latter time can be estimated as
\beq
\tau_{BH} \sim \exp(2\pi/H) \sim \exp(10^{62}),
\eeq
where the numerical estimate is for the observed value of $H\sim 10^{-61}$.  Clearly, the condition $\tau_{obs}<\tau_{BH}$ is satisfied with a very wide margin, and it seems likely that it is satisfied for all anthropic vacua (which require small values of $H$ to allow structure formation).

\subsection{Other measures}

The issue of black hole nucleation and its effect on the measure arises not only for the watcher measure, but for other measure proposals as well.  In all local measure prescriptions, we are instructed to follow a timelike geodesic, so some rule needs to be specified what should be done when the geodesic encounters a black hole.  All global measures utilize a congruence of timelike geodesics, so once again we need instructions on what to do when geodesics encounter black holes.  Note that this issue is unrelated to whether or not singularities are resolved at AdS crunches and in black hole interiors.  If black hole singularities are not resolved, then geodesics must be terminated at singularities, but still this has a significant effect on the measure.

For local measures, this problem can be addressed along the same lines as we discussed for the watcher measure in the preceding subsection.  For global measures, if one ignores geodesics captured by black holes, it is not clear how one should deal with the holes that will as a result develop in the congruence.  The holes may be partially closed as the geodesics are deflected in the gravitational field of black holes.  They could also be fully closed, in which case the congruence will develop caustics after passing the black hole.  





\section{Summary and discussion}

We have reconsidered the measure problem of inflationary cosmology, by introducing the non-standard assumption that
spacetime singularities are resolved in the fundamental theory, in such a way that all time-like geodesics can be extended
indefinitely into the future. This allows us to define a measure based on a single future-eternal time-like geodesic. This geodesic 
can be thought of as the world-line of a ``watcher", sampling different types of events as they are intersected in the course of time.
An immediate consequence of this approach is that the measure is independent of initial conditions, due to the attractor behaviour 
of the rate equations determining the frequencies at which the different types of events are sampled.

Aside from the dependence on initial conditions, previous versions of geodesic-based measures suffer from ambiguities associated with
the choice of a sampling cut-off region in the vicinity of the geodesic. Here, we circumvent this ambiguity by counting any ``stories" which are
pierced through by the watcher's geodesic. This avoids any reference to cut-offs. Size bias is eliminated by weighing each occurrence by the inverse of the cross-section of the story under consideration. 

Phenomenologically, this measure is quite similar to the fat geodesic measure. 
Since the fat geodesic measure does not suffer from any obvious phenomenological problems, we expect the watcher's measure to do just as well. 

A major difference between the present approach and the standard picture of the multiverse is that the transitions occurring at the bounces (which replace the would-be singularities), are expected to 
lead to significant violations of detailed balance. As a result, the
fraction $f_j$ of time spent by the watcher in the different inflating
vacua $j$ in the landscape can be very far from ergodic. This feature
is welcome, since exact ergodicity entails thermal death and Boltzmann
brain dominance. In the standard case, where bounces are not allowed,
the presence of Minkowski and AdS terminal vacua can still generate
significant departures from ergodicity. However, it is unclear that
this is sufficient in order to eliminate the problem of BB dominance
in generic landscapes \cite{DGLNSV,Bousso:2008hz,GGV12}. Here, we have
argued that the effect of bounces can alleviate the BB problem,
especially if bounces can also occur in black hole interiors.  The
watcher's geodesic typically encounters a (mini) black hole a
relatively short time after crossing into a new bubble, so the time
available for encountering BBs is reduced compared to the standard
scenario. 

In the absence of bounces, the nature of the distribution $f_j$ in the
string theory landscape  was discussed in
Ref.~\cite{Delia}. There, it was argued that the distribution
is strongly peaked at the `dominant' dS vacuum $D$, which has the
slowest decay rate and is likely to have a very small energy density.
The values of $f_j$ for all anthropic vacua are then suppressed by
extremely small upward transition rates $\Gamma_{jD}$ from $D$.  The
phenomenological implications of this picture have been recently
discussed in Refs.~\cite{Douglas,Susskind12}, where it is argued that
it suggests a low-energy supersymmetry breaking and a low energy scale
of inflation. In the presence of AdS bounces, the dominant vacuum
picture does not apply, and the conclusions of
\cite{Douglas,Susskind12} no longer hold.  A more detailed discussion
of the phenomenology of the watcher measure will be given elsewhere. 

{In the Introduction, we mentioned that the holographic approach to the measure problem \cite{GVhm,GVhmci}
encounters difficulties in the presence of AdS crunches, since it is not clear what to use as a holographic screen for the
interior of AdS bubbles. On the other hand, if we assume that crunches are substituted by bounces, then the 
structure of the future boundary of the multiverse is modified. Whether this non-singular future boundary can be used for a holographic 
description remains an open question which we leave for further study.}

After this work was completed, we became aware of the papers by {Piao \cite{P1,P2}, where the picture of a multiverse with AdS bounces was
first discussed}.

We also became aware of the papers by Johnson and
Lehners  \cite{JL} and by Lehners \cite{J-L}, where some models of eternal inflation are
studied allowing for the possibility of non-singular bounces.
Specifically, they investigated scenarios where ekpyrosis or cyclic
universes occur inside some of the bubbles. In these scenarios,
the bounce is supposed to lead to a hot phase followed by a dark energy dominated phase, 
at a very low energy scale. This is in contrast with the approach we are adopting here,
where we argue that generic crunches, such as AdS crunches or the singularities inside of black holes, 
may also lead to transitions to high energy inflating vacua. It would be interesting to reconsider the scenarios 
in Refs. \cite{JL,J-L} within the present approach.

\subsection*{Acknowledgements}

We are grateful to Alan Guth, Ken Olum, Raphael Bousso and Matt Kleban for stimulating discussions.
This work was made possible by grants from the Templeton Foundation, PHY-0855447 from the National Science 
Foundation, AGAUR 2009-SGR-168, MEC FPA 2010-20807-C02- 02 and CPAN CSD2007-00042 Consolider-Ingenio 2010. 
J.G. thanks the Tufts Cosmology Institute and the Yukawa Institute for Theoretical Physics 
for hospitality during the preparation of this work.

\section{Appendix A}

The relevant properties of the transition matrices $T$ and $M$ can be
deduced from the Perron-Frobenius theorem.  The theorem can be stated as follows \cite{Perron1,Perron2}: 

An irreducible matrix $A_{IJ}$ with non-negative elements, $A_{IJ}\geq 0$, has a positive, non-degenerate eigenvalue $\gamma_0\geq 0$ such that all other eigenvalues $\gamma_a$ satisfy 
\beq
|\gamma_a|<\gamma_0.  
\label{gammaa}
\eeq
The eigenvector corresponding to $\gamma_0$ can be chosen with all positive components.  Furthermore, if we denote
\beq
\sigma_J\equiv\sum_I A_{IJ} ,
\eeq
then $\gamma_0$ is bounded by
\beq
{\rm min}_J \sigma_J \leq \gamma_0 \leq {\rm max}_J \sigma_J.
\label{PF}
\eeq

In our case, the matrix $T_{IJ}$ is irreducible and $T_{IJ}\geq 0$; hence the Perron-Frobenius theorem applies.  Moreover, from Eq.~(\ref{unitarity}), 
\beq
\sigma_J = \sum_I T_{IJ} = 1
\eeq
for all $J$, and it follows from (\ref{PF}) that $\gamma_0=1$.  Since this eigenvalue is nondegenerate and all other eigenvalues satisfy (\ref{gammaa}), it follows that
\beq
\Re \gamma_a < \gamma_0.
\eeq

The matrix $M$ in Eq.~(\ref{M}) is irreducible and has non-negative off-diagonal elements, $M_{ij}\geq 0$ for $i\neq j$, but he diagonal elements satisfy $M_{ii}\leq 0$.  So the Perron-Frobenius theorem does not directly apply to $M$, but it can be applied to the matrix ${\tilde M} = M - \zeta I$, where $\zeta = \min M_{ii}$ and $I$ is the unit matrix.    Since
\beq
\sum_i M_{ij} = 0,
\eeq
we have
\beq
{\tilde\sigma}_j \equiv \sum_i {\tilde M}_{ij} =-\zeta ,
\eeq
and it follows from (\ref{PF}) that the Perron-Frobenius eigenvalue of ${\tilde M}$ is ${\tilde \gamma}_0 = -\zeta \geq 0$.  And it follows immediately that the dominant (having the largest real part) eigenvalue of $M$ is $\beta=0$.

\end{document}